\documentclass[prb,aps,twocolumn,footinbib,floatfix,10pt,longbibliography,superscriptaddress,notitlepage]{revtex4-1}
\usepackage[english]{babel}
\usepackage{breakcites}
\usepackage{listings}
\usepackage[utf8]{inputenc}
\usepackage[T1]{fontenc}
\usepackage{textcomp}
\usepackage{gensymb,braket}
\pdfoutput=1
\usepackage{graphicx}
\usepackage{braket}
\usepackage{color}
\usepackage{epstopdf}
\usepackage{mathtools}
\usepackage{amsmath}
\usepackage{amssymb}
\usepackage{float}
\usepackage{comment}
\usepackage{amsmath}

\usepackage{amsfonts}
\usepackage{bm}
\usepackage{bbold}

\usepackage{color}

\graphicspath{{./figs/}}
\usepackage{gensymb}
\usepackage[colorinlistoftodos,prependcaption]{todonotes}
\usepackage{xargs}  
\newcommandx{\unsure}[2][1=]{\todo[linecolor=red,backgroundcolor=red!25,bordercolor=red,#1]{#2}}
\newcommandx{\change}[2][1=]{\todo[linecolor=blue,backgroundcolor=blue!25,bordercolor=blue,#1]{#2}}
\newcommandx{\info}[2][1=]{\todo[linecolor=OliveGreen,backgroundcolor=OliveGreen!25,bordercolor=OliveGreen,#1]{#2}}
\newcommandx{\improvement}[2][1=]{\todo[linecolor=Plum,backgroundcolor=Plum!25,bordercolor=Plum,#1]{#2}}
\newcommandx{\thiswillnotshow}[2][1=]{\todo[disable,#1]{#2}}
\newcommandx{\greencom}[2][1=]
{\todo[inline, color=green!40,#1]{#2}}
\newcommandx{\bluecom}[2][1=]
{\todo[inline, color=blue!40,#1]{#2}}

\usepackage[colorlinks]{hyperref}
\definecolor{winered}{rgb}{0.5,0,0}
\hypersetup
{
colorlinks=true,
linkcolor=winered,
urlcolor={winered},
filecolor={winered},
citecolor={winered},
allcolors={winered}
}

\usepackage{letltxmacro}
\LetLtxMacro{\ORIGselectlanguage}{\selectlanguage}
\makeatletter
\DeclareRobustCommand{\selectlanguage}[1]{%
  \@ifundefined{alias@\string#1}
    {\ORIGselectlanguage{#1}}
    {\begingroup\edef\x{\endgroup
       \noexpand\ORIGselectlanguage{\@nameuse{alias@#1}}}\x}%
}
\newcommand{\definelanguagealias}[2]{%
  \@namedef{alias@#1}{#2}%
}
\makeatother

\definelanguagealias{en}{english}
\definelanguagealias{EN}{english}

\graphicspath{{figure/}}

\begin{document}
\title{
Fermi's golden rule for  spontaneous emission in absorptive and amplifying  media}.
\author{Sebastian Franke}
\email{sebastian.franke@tu-berlin.de}
\affiliation{Technische Universit\"at Berlin, Institut f\"ur Theoretische Physik,
Nichtlineare Optik und Quantenelektronik, Hardenbergstra{\ss}e 36, 10623 Berlin, Germany}
\affiliation{\hspace{0pt}Department of Physics, Engineering Physics, and Astronomy, Queen's University, Kingston, Ontario K7L 3N6, Canada\hspace{0pt}}
\author{Juanjuan Ren}
\affiliation{\hspace{0pt}Department of Physics, Engineering Physics, and Astronomy, Queen's University, Kingston, Ontario K7L 3N6, Canada\hspace{0pt}}
\author{Marten Richter}
\affiliation{Technische Universit\"at Berlin, Institut f\"ur Theoretische Physik,
 Nichtlineare Optik und Quantenelektronik, Hardenbergstra{\ss}e 36, 10623 Berlin, Germany}
   \author{Andreas Knorr}
 \affiliation{Technische Universit\"at Berlin, Institut f\"ur Theoretische Physik,
 Nichtlineare Optik und Quantenelektronik, Hardenbergstra{\ss}e 36, 10623 Berlin, Germany}
 \author{Stephen Hughes}
\affiliation{\hspace{0pt}Department of Physics, Engineering Physics, and Astronomy, Queen's University, Kingston, Ontario K7L 3N6, Canada\hspace{0pt}}

\date{\today}

\begin{abstract}
 We demonstrate a fundamental breakdown of
the 
photonic spontaneous emission (SE) formula derived from Fermi's golden rule,  in absorptive and amplifying media, where one assumes the
SE rate scales with the local
photon density of states, an approach often used in more complex, semiclassical nanophotonics simulations.
Using a rigorous quantization of the macroscopic Maxwell equations in the presence of arbitrary linear media, we derive a corrected Fermi's golden rule and master equation for a quantum 
two-level system (TLS) that yields
a quantum pumping term and a modified decay rate that is net positive.
We show rigorous numerical results of the temporal dynamics of the TLS for an 
example of two coupled  microdisk resonators, 
forming a gain-loss medium,
 and demonstrate the clear failure of the  
commonly adopted formulas based solely on the local density of states.

\end{abstract}
\maketitle 

Spontaneous emission (SE) of a two-level system (TLS) has been a fundamental topic since the birth of quantum electrodynamics~\cite{dirac1927quantum,Weisskopf1930,RevModPhys.4.87} and is one of the standard metrics for many applications in nanophotonic and nanoplasmonic systems 
interacting with quantum emitters, such as molecules and quantum dots. 
Photonic engineering allows one to modify the
local density of states (LDOS) which in turn leads to an enhancement or a suppression of the SE decay rate~\cite{PhysRevLett.58.2059,Lodahl2004},
 $\Gamma$, which is associated with the transition probabilities with respect to a pertubation $\hat{V}$, formalized into Fermi's golden rule (FGR). 
\begin{figure}[t]
 \centering
 \includegraphics[width=1\columnwidth,trim=0 0.0cm 0 0.2cm, clip]{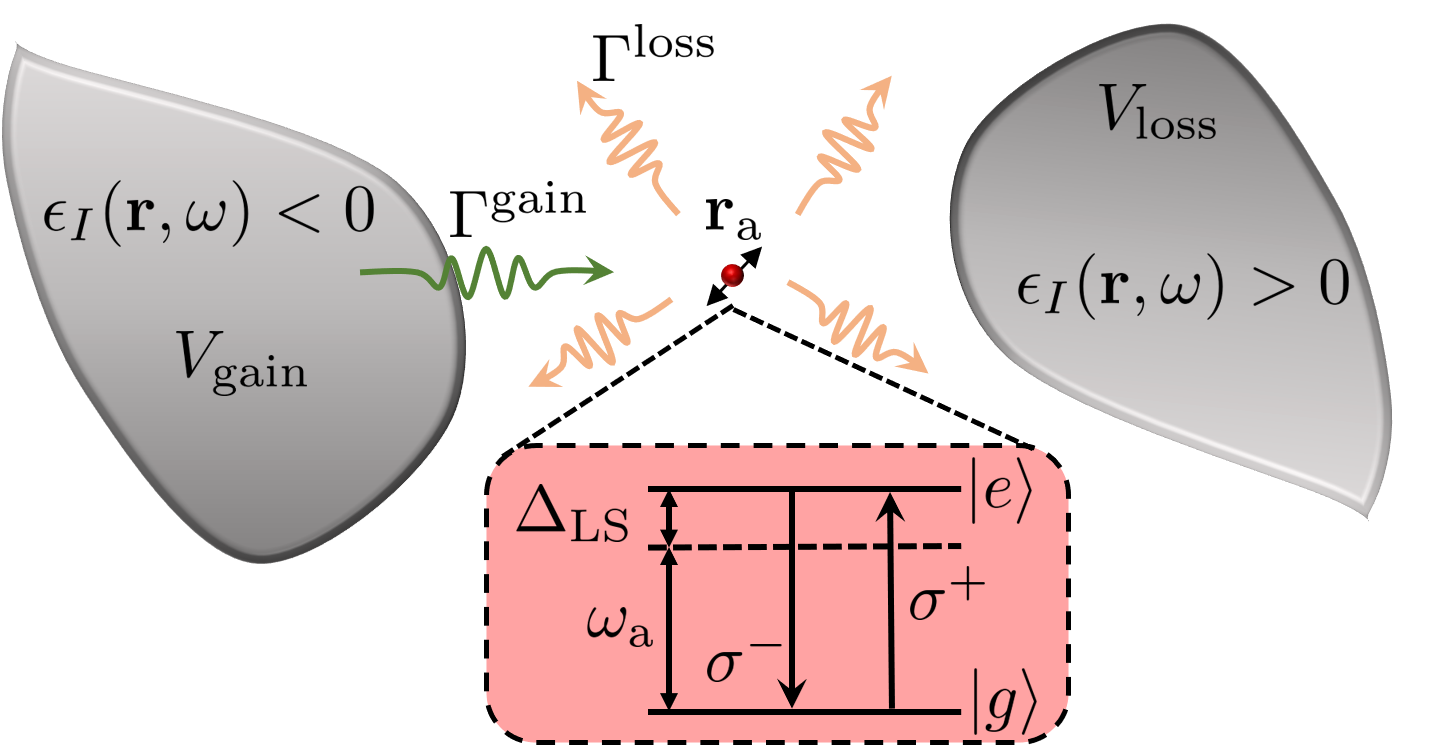}
 \caption{{Schematic of a quantum emitter (red dot) at position $\mathbf{r}_{\rm a}$, treated as a 
 TLS 
 and weakly interacting with an arbitrary amplifying ($\epsilon_{I}(\mathbf{r},\omega)<0$) and lossy ($\epsilon_{I}(\mathbf{r},\omega)>0$) 
 photonic 
 environment (grey areas). The emitter is pumped by the amplifying medium with a rate $\Gamma^{\rm gain}$ and its free spontaneous decay $\Gamma_0$ is enhanced by $\Gamma^{\rm loss}/\Gamma_0$.
 The environment also induces a  gain-loss dependent photonic Lamb shift $\Delta_{\rm LS}$.}}
\label{fig: SchematicIntro}
\end{figure}

In terms of a TLS 
quantum 
dipole interaction and the photonic LDOS, 
the 
 FGR in the photonics community is
\begin{equation}
    \Gamma^{\rm LDOS}({\bf r}_{\rm a})=\frac{2}{\hbar\epsilon_0} {\bf d} \cdot {\rm Im}[{\bf G}({\bf r}_{\rm a}, {\bf r}_{\rm a},\omega_{\rm a})] \cdot {\bf d}^*\label{eq: FGR_class},
\end{equation}
where $\mathbf{G}$ is the photonic Green function at the position $\mathbf{r}_{\rm a}$ of the TLS, with a 
 dipole moment 
 ${\bf d} =\langle e|\hat{\bf d}|g\rangle=d\mathbf{n}_{\rm d}$
 and frequency $\omega_{\rm a}$. 
From a quantum field theory,
the SE rate is also connected to the {\it vacuum fluctuations}~\cite{Milonni1976} of the quantized electric field ($\hat {\bf E}$), 
through 
\begin{equation}
    \Gamma^{\rm VF}({\bf r}_{\rm a})
    =\frac{2\pi}{\hbar^2}\mathbf{d}\cdot\langle 0|[\hat{\mathbf{E}}(\mathbf{r}_{\rm a},\omega_{\rm a}),\hat{\mathbf{E}}^\dagger(\mathbf{r}_{\rm a},\omega_{\rm a}) ]|0\rangle\cdot\mathbf{d}^*,\label{eq: SEClass2}
\end{equation}
 where $|0\rangle$ is the vacuum state.
The interpretation of SE decay, 
on a quantum level, is a consequence of radiation reaction 
and 
vacuum fluctuations~\cite{Milonni1976,PhysRevLett.77.2444,dung2000spontaneous, PhysRevA.61.033807}, which are identical in the case of lossy media, where both simply depend on the LDOS. Thus, in purely lossy environments, the theory of SE decay can often be developed equally from a classical treatment of light and matter through radiation reaction terms from Poynting's theorem~\cite{Hecht}, and from a weak coupling limit of quantized light-matter interactions through vacuum fluctuations and radiation reaction terms. 
In the following, such approaches utilizing Eq.~\eqref{eq: FGR_class} for calculating SE are denoted as LDOS-SE models.

Consequently, 
Eqs.~\eqref{eq: FGR_class} and \eqref{eq: SEClass2} are commonly used as a basis for calculating important 
figures of merit, such as  the Purcell factor and  radiative  $\beta$-factor, even with 
gain media~\cite{PhysRevLett.117.107402,Pick:17,PhysRevB.102.155303,
PhysRevA.64.033812}. Indeed, 
Purcell's formula~\cite{purcell1946resonance} is simply a special case of Eqs.~\eqref{eq: FGR_class}-\eqref{eq: SEClass2} for a single electromagnetic mode, aligned to the dipole and frequency of the TLS.

In this Letter, we show 
that this common LDOS view of SE~\cite{Pick:17,PhysRevB.102.155303,
PhysRevA.64.033812,PhysRevA.99.033853,LiuLeiWangLongWang} is in general {\it incorrect}, 
and present a 
revision to 
the usual FGR 
for the SE decay
of a point dipole (correcting Eq.~\eqref{eq: FGR_class}), 
for the case of dielectric media with amplifying regions.
From a macroscopic quantum theory of loss and gain, 
we derive the 
master equation of a TLS interacting with its photonic environment in the weak coupling limit, through a second-order Born-Markov approximation, which gives rise to terms associated with the pump (from the gain region) and the decay (from the loss region) of the emitter. 
Using a coupled gain-loss resonator structure, we show that while the LDOS-SE model leads to negative Purcell factors and nonphysical temporal evolutions of the TLS, our scheme leads to a {\em positive Purcell factor with non-vanishing steady states}.
We explain why the common assumption that the SE rate scales with the LDOS  {\em fails}, and 
show that 
the SE in such media must be derived from a proper quantum model, taking into account the ordering of the field operators that  have more significant consequences compared to the lossy medium case.

\textit{Theory.}
We  consider 
the general Hamiltonian $H=H_{\rm a}
+H_B + H_I$ of a TLS interacting with the electromagnetic field in a  lossy~\cite{Dung,dung2000spontaneous} and amplifying media~\cite{raabe2008qed}, with
\begin{subequations}
\label{eq: Htotal}
\begin{align}
    H_{\rm a}&=\hbar\omega_{\rm a}\sigma^+\sigma^-,\\
    H_B &= \hbar\int{\rm d}\mathbf{r}~{\rm sgn}(\epsilon_I)\int_0^\infty{\rm d}\omega~\omega \mathbf{b}^\dagger(\mathbf{r},\omega)\cdot\mathbf{b}(\mathbf{r},\omega),\\
    H_I &= -\left[\sigma^+\int_0^\infty{\rm d}\omega \mathbf{d}\cdot\hat{\mathbf{E}}(\mathbf{r}_{\rm a},\omega)+{\rm H.a.}\right],
\end{align}
\end{subequations}
where the spatial integral is over all 
space,  ${\rm sgn}$ is the sign function, $\epsilon_{I}$ is the imaginary part of the permittivity, 
 $\sigma^{\pm}$ are the Pauli operators, and
$\mathbf{b}^{(\dagger)}(\mathbf{r},\omega)$ are the bosonic annihilation (creation) operators of the medium and the electromagnetic degrees of freedom. 
The medium-assisted electric field operator $\hat{\mathbf{E}}(\mathbf{r}_{\rm a},\omega)$
fulfills the Helmholtz equation,
    $\left[\boldsymbol{\nabla}\times\boldsymbol{\nabla}\times-\epsilon(\mathbf{r},\omega)\omega^2/c^2\right]\hat{\mathbf{E}}(\mathbf{r},\omega)=i\omega\mu_0\hat{\mathbf{j}}_{\rm N}(\mathbf{r},\omega)$,
where $ \hat{\mathbf{j}}_{\rm N}(\mathbf{r},\omega)$ is the current noise operator, which preserves the fundamental QED commutation relation for arbitrary media~\cite{franke2020fluctuation}; In the presence of lossy and amplifying media~\cite{raabe2008qed}, then $\hat{\mathbf{j}}_{\rm N}(\mathbf{r},\omega){=}\omega\sqrt{\hbar\epsilon_0|\epsilon_I(\mathbf{r},\omega)|/\pi}[\Theta(\epsilon_I)\hat{\mathbf{b}}(\mathbf{r},\omega)+\Theta(-\epsilon_I)\hat{\mathbf{b}}^\dagger(\mathbf{r},\omega)]$, 
where $\Theta[\epsilon_I]$ ($\Theta[-\epsilon_I]$) is the Heaviside function with respect to the spatial region, $\mathbb{R}^3-V_{\rm gain}$ ($V_{\rm gain}$), with passive (active) dielectric permittivity $\epsilon_I(\mathbf{r},\omega)>0$ ($\epsilon_I(\mathbf{r},\omega)<0$), 
cf.~Fig.~\ref{fig: SchematicIntro}. The introduction of a phenomenological noise operator in  purely lossy (as well as in lossy and amplifying) media 
is rigorously justified 
using a  microscopic oscillator model~\cite{Suttorp,philbin2010canonical,PhysRevA.84.013806}. We also provide a short overview of the development of the corresponding literature in the Supplemental Material~\cite{SI_short}.

The Helmholtz equation has the 
source-field 
solution $\hat{\mathbf{E}}(\mathbf{r},\omega){=}i\int{\mathrm d}{\bf s} \mathbf{G}(\mathbf{r},\mathbf{s},\omega){\cdot} \hat{\mathbf{j}}_{\rm N}(\mathbf{s},\omega)/(\omega\epsilon_0)$, where $\mathbf{G}(\mathbf{r},\mathbf{s},\omega)$ is the
Green function of the medium, which 
satisfies
$[\boldsymbol{\nabla}{\times}\boldsymbol{\nabla}{\times}
{-}\epsilon(\mathbf{r},\omega)\omega^2/c^2]\mathbf{G}(\mathbf{r},\mathbf{s},\omega){=}\omega^2\mathbb{1}\delta(\mathbf{{r}-\mathbf{s}})/c^2$,~with suitable radiation boundary conditions. 
The macroscopic Green function quantization is valid under strict linear response of the electromagnetic field in the dielectric medium~\cite{raabe2008qed} (cf. Ref.~\onlinecite{SI_short}).

\begin{figure*}
  \includegraphics[width=1\columnwidth]{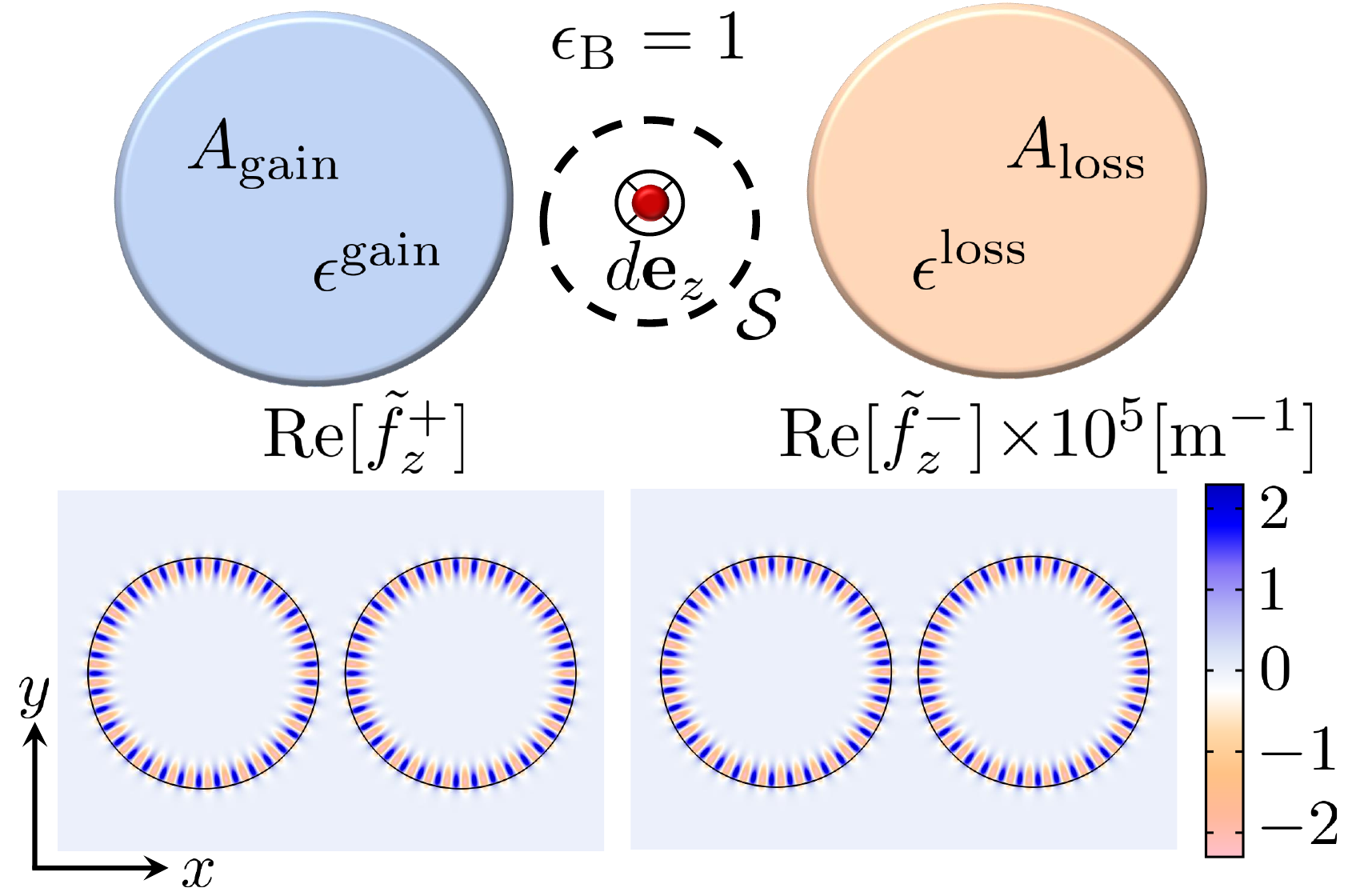}
   \includegraphics[trim={0.25cm 0.25cm 0.25cm 0.1cm},clip,width=1\columnwidth]{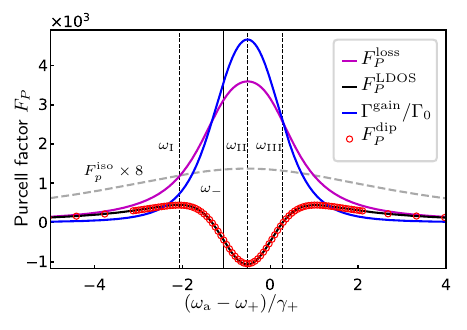}
   \caption{Left: Schematic of a $z$-polarized emitter with dipole moment $d$
   placed  between a lossy and amplifying
   microdisk resonator, as well as the two hybridized QNMs (2D model, areas $A^{\rm gain/loss}$ with disk radius
   $R{=}5\,\mu$m and gap distance $d_{\rm gap}{=}1155~$nm). The dielectric constants of the resonators are 
   $\epsilon^{\rm loss}{=}(2{+}10^{-5}i)^2 $ and $\epsilon^{\rm gain}{=}(2{-}5\cdot 10^{-6}i)^2$, giving rise to nearly degenerate QNM eigenfrequencies $\hbar\tilde{\omega}_+ [{\rm eV}]{=}0.833717 {-}1.021253 \cdot 10^{-6}i$ and $\hbar\tilde{\omega}_-[{\rm eV}]= 0.833716-1.038998\cdot 10^{-6}i $ (cf. Ref.~\onlinecite{EPClassicalPaper} for numerical details on the QNM calculations). Right: Purcell factor of the emitter-cavity system as function of TLS frequency shifted with respect to real part of the QNM hybrid frequencies $\tilde{\omega}_+$, showing results obtained from LDOS-SE model for the isolated lossy disk (grey dashed) and hybrid structure (black), Eq.~\eqref{eq: FGR_class}, the full numerical Maxwell-dipole solution (red circles), 
   as well as the rigorous FGR (magenta, Eq.~\eqref{eq: GammaLoss}). Additionally, the contribution of the pump from FGR is shown (blue). 
   }
   \label{fig: Result1}
\end{figure*}

\textit{Breakdown of the LDOS-SE rate}.\
To derive the correct SE  for amplifying media, we consider the time derivative of the density operator $\langle f,\mathbf{1}|\rho |f,\mathbf{1}\rangle$, with a final TLS state $|f\rangle$ at time $t$ with respect to an initial TLS state $|i,\mathbf{0}\rangle$ ($|f\rangle\neq|i\rangle$) at time $t_0$. Treating the interaction between the TLS and the photons pertubatively up to the second order leads to FGR, and we derive the transition rate from $|e,0\rangle$ to $|g,\mathbf{1}\rangle$ as $\Gamma_{(e,0)\rightarrow (g,\mathbf{1})}\equiv \Gamma^{\rm loss}$,  where~\cite{SI_short}
\begin{equation}
   \Gamma^{\rm loss}=\frac{2}{\hbar\epsilon_0} {\bf d} \cdot \mathbf{K}(\mathbf{r}_{\rm a},\mathbf{r}_{\rm a},\omega_{\rm a}) \cdot {\bf d}^*,\label{eq: GammaLoss}
\end{equation}
where $\mathbf{K}(\mathbf{r},\mathbf{r}')=\int_{\mathbb{R}^3-V_{\rm gain}}{\rm d}\mathbf{s}\epsilon_I(\mathbf{s})\mathbf{G}(\mathbf{r},\mathbf{s})\cdot\mathbf{G}^*(\mathbf{s},\mathbf{r}')$; notably, in the purely lossy case ($V_{\rm gain}\rightarrow \emptyset$), this leads to the LDOS-SE formula, Eq.~\eqref{eq: FGR_class}, through the Green identity~\cite{Dung,franke2020fluctuation} $\int_{\mathbb{R}^3}{\rm d}\mathbf{s}\epsilon_I(\mathbf{s})\mathbf{G}(\mathbf{r},\mathbf{s})\cdot\mathbf{G}^*(\mathbf{s},\mathbf{r}')={\rm Im}[\mathbf{G}(\mathbf{r},\mathbf{r}')]$.
Similarly, the transition rate from $|g,0\rangle$ to $|e,\mathbf{1}\rangle$ is obtained, $\Gamma_{(g,0)\rightarrow (e,\mathbf{1})}\equiv \Gamma^{\rm gain}$, where
\begin{equation}
     \Gamma^{\rm gain}(\mathbf{r}_{\rm a})=\frac{2}{\hbar\epsilon_0} {\bf d} \cdot \mathbf{I}(\mathbf{r}_{\rm a},\mathbf{r}_{\rm a},\omega_{\rm a}) \cdot {\bf d}^*,\label{eq: GammaGain}
\end{equation}
and $\mathbf{I}(\mathbf{r},\mathbf{r}')=\int_{V_{\rm gain}}{\rm d}\mathbf{s}|\epsilon_I(\mathbf{s})|\mathbf{G}(\mathbf{r},\mathbf{s})\cdot\mathbf{G}^*(\mathbf{s},\mathbf{r}')$. Note 
that $\Gamma^{\rm gain}({\bf r}_{\rm a})\geq 0$, since $|\epsilon_I|\geq 0$ and the integral can be recast into a form $\int |\mathbf{f}|^2$ by using properties of the outer product and the reciprocity theorem, $\mathbf{G}(\mathbf{r},\mathbf{r}')=\mathbf{G}^{\rm T}(\mathbf{r}',\mathbf{r})$. This same holds true for $\Gamma^{\rm loss}$.

First, we recognize  that the form of $\Gamma^{\rm gain}({\bf r}_{\rm a})$  and $\Gamma^{\rm loss}({\bf r}_{\rm a})$ is fundamentally different from the usually assumed LDOS-SE formula in the photonics community; however, we can relate them to Eq.~\eqref{eq: FGR_class}, so that
\begin{equation}
    \Gamma^{\rm loss}({\bf r}_{\rm a})=\Gamma^{\rm LDOS}(\mathbf{r}_{\rm a})+\Gamma^{\rm gain}({\bf r}_{\rm a}),
\end{equation}
and from 
Eqs.~\eqref{eq: FGR_class} and \eqref{eq: GammaLoss}, we deduce that
$\Gamma^{\rm LDOS}=\Gamma^{\rm loss}-\Gamma^{\rm gain}$, which implies that the sole use of $\Gamma^{\mathrm{LDOS}}$ is
{\textit{clearly inconsistent}} for SE with the rigorous treatment of FGR in the presence of gain. 
Second, we can investigate the contributions associated with vacuum fluctuations.
Inserting the source field expansion of the electric field into Eq.~\eqref{eq: SEClass2}, then
 $\Gamma^{\rm VF}=\Gamma^{\rm LDOS}$~\cite{SI_short}, identical to the results obtained from the commonly used FGR;
 however, this is only
because the more general commutator instead of the anti-normal ordered operator expression is used as the basis for the quantum (vacuum) fluctuations, Eq.~\eqref{eq: SEClass2}. 
In fact, compared to the purely lossy case, the normal ordered contribution to the electric field commutator, $\langle 0|\hat{\mathbf{E}}^\dagger(\mathbf{r}_{\rm a},\omega_{\rm a})\hat{\mathbf{E}}(\mathbf{r}_{\rm a},\omega_{\rm a})|0\rangle$, does not vanish. In the rigorous formulation of FGR, this term is precisely related to the gain rate $\Gamma^{\rm gain}$ (cf. Ref.~\onlinecite{SI_short}).
Thus, the LDOS is still related to quantum (vacuum) fluctuations, but it is no longer related to a physically relevant transition rate in the presence of gain. 
Note, that from the spatial integral expressions of $\Gamma^{\mathrm{gain}}$ and $\Gamma^{\mathrm{loss}}$, one can clearly see, that $\Gamma^\textrm{LDOS}$ as a difference between two positive definite forms is in general not positive definite.

\textit{Dynamical equations of TLS densities and optical dephasing}.\
 To obtain a more general description of the TLS dynamics in the presence of gain, we use the density matrix picture, starting with the Liouville-von Neumann equation $\partial_t\rho=-i[H,\rho]/\hbar$, where $\rho$ is the total density operator and $H$ is Hamiltonian from Eq.~\eqref{eq: Htotal}. Applying the second-order Born Markov approximation, we obtain the reduced TLS master equation, whose projections on the basis elements of the TLS subspace are~\cite{SI_short}
 \begin{subequations}\label{eq: rhoDynamics}
 \begin{align}
    \partial_t\rho_{\rm a}^{ee}=-\Gamma^{\rm loss}\rho_{\rm a}^{ee}+\Gamma^{\rm gain}\rho_{\rm a}^{gg},~\partial_t\rho_{\rm a}^{gg}=-\partial_t\rho_{\rm a}^{ee}\label{eq: rho_ee_gg},\\
    \partial_t\rho_{\rm a}^{eg}=-i[\omega_{\rm a}+\Delta_{\rm LS}]\rho_{\rm a}^{eg}-[\Gamma^{\rm loss}+\Gamma^{\rm gain}]\rho_{\rm a}^{eg}\label{eq: rho_eg}, 
\end{align}
 \end{subequations}
 where $\rho_{\rm a}={\rm tr}_{\rm B}\rho$ is the trace with respect to the electromagnetic degrees of freedom, $\rho_{\rm a}^{ee}=\langle e|\rho_{\rm a}|e\rangle$ ($\rho_{\rm a}^{gg}=\langle g|\rho_{\rm a}|g\rangle$) is the excited (ground) state density and $\rho_{\rm a}^{eg}=\langle g|\rho_{\rm a}|e\rangle$ is the dephasing with Lamb shift $\Delta_{\rm LS}=\mathbf{d}\cdot{\rm Re}[\mathbf{G}(\mathbf{r}_{\rm a},\mathbf{r}_{\rm a},\omega_{\rm a})]\cdot \mathbf{d}/(\hbar\epsilon_0)$. Compared to the LDOS-SE model, the ground state and excited state densities are coupled through the loss and gain rate. Moreover, the dephasing decays with $\Gamma^{\rm loss}+\Gamma^{\rm gain}$ rather then $\Gamma^{\rm loss}-\Gamma^{\rm gain}$.  Interestingly, the effective Lamb shift $\Delta_{\rm LS}$ of the dephasing oscillation is identical to the purely lossy case.

Consequently, {\em we find three fundamental corrections to common formulas in the current literature}: (i) the rate obtained from the LDOS-SE model
is corrected by an additional term $-\Gamma^{\rm gain}$
; (ii) a non-vanishing transition process from the equilibrium ground  state $|g,0\rangle$ to the excited states with one gain media excitations missing $|e,\mathbf{1}_{\rm gain}\rangle$ occurs; (iii) the gain and loss rate appear as coupling constants in the dynamical equations of the TLS densities, which is clearly missing in the usual models~\cite{PhysRevLett.117.107402,Pick:17,PhysRevB.102.155303,
PhysRevA.64.033812,PhysRevA.99.033853}, which associate all interaction processes as decay terms. Thus, Eqs.~\eqref{eq: rhoDynamics}(a-b) corrects the
pertubative limit expression
of a quantum emitter in the presence of amplifying and lossy media.

It is also interesting to investigate the classical limit of the underlying master equation of Eqs.~\eqref{eq: rhoDynamics}(a-b). This is achieved by treating the TLS as a (bosonic) harmonic oscillator through $[\sigma^-,\sigma^+]\rightarrow 1$;
then 
the excited state density 
evolves via~\cite{SI_short} $\partial_t\rho_{\rm a}^{ee}(t) = -\Gamma^{\rm LDOS}\rho_{\rm a}^{ee}(t)+\Gamma^{\rm gain}$, which is identical to the behavior of the LDOS-SE model apart from the constant pumping term $\Gamma^{\rm gain}$. However, this assumption is only valid in the weak excitation limit, where $\Gamma^{\rm gain}\ll \Gamma^{\rm loss}$. Indeed, for any case where amplification constitutes a significant part of the dielectric medium, the classical limit of the TLS is violated, and the use of the corrected quantum model is necessary. 

\textit{Numerical results for coupled microdisk resonators}.
To explicitly demonstrate the impact and consequences of the corrected formula for SE, we now investigate the Purcell enhancement and dynamics of a TLS in an examplary cavity-QED setup, where the quantum dipole is placed in the gap between a lossy and an amplifying resonator, as illustrated in Fig.~\ref{fig: Result1}.
Such resonators are commonly used
to explore the enhanced sensing capabilities, lasing, and unidirectional transmission near exceptional points~\cite{peng_loss-induced_2014,peng_paritytime-symmetric_2014,chang_paritytime_2014,chen_exceptional_2017,chen_parity-time-symmetric_2018,miri_exceptional_2019}.

For calculating the quantum parameters appearing in Eq.~\eqref{eq: rhoDynamics}, the classical photon Green function must be determined, which can be a tedious task for arbitrary shaped scattering structures. However, it has been shown 
that for dipole positions $\mathbf{r}_{\rm a}$ and a reference position $\mathbf{r}$ near or inside the cavity regions, then the scattering part of the Green function can be accurately represented by a quasinormal mode (QNM) expansion of the form~\cite{MDR1,2ndquant2,Kristensen:20,PhysRevLett.122.213901,EPClassicalPaper} 
\begin{equation}
    \mathbf{G}_{\rm ff}(\mathbf{r}_{\rm a},\mathbf{r},\omega)=\sum_{\mu} A_\mu(\omega)\tilde{\mathbf{f}}_\mu(\mathbf{r}_{\rm a})\tilde{\mathbf{f}}_\mu(\mathbf{r}),\label{eq: GFQNM}
\end{equation}
where $A_\mu(\omega){=}\omega/(2(\tilde{\omega}_\mu{-}\omega))$ is the particular choice of the spectral coefficient,  $\tilde{\omega}_\mu$ are the QNM  eigenfrequencies, and $\tilde{\mathbf{f}}_\mu(\mathbf{r})$ is the QNM eigenfunction, solving $\left[\boldsymbol{\nabla}\times\boldsymbol{\nabla}\times-\epsilon(\mathbf{r},\tilde{\omega}_\mu)\tilde{\omega}_\mu^2/c^2\right]\tilde{\mathbf{f}}_\mu(\mathbf{r}){=}0$ together with open boundary conditions. Due to the choice of outgoing radiation conditions, the QNM eigenfrequencies are complex with $\tilde{\omega}_\mu = \omega_\mu - i\gamma_\mu$, where $\gamma_\mu$ is the HWHM of the QNM resonance with center frequency $\omega_\mu$. For $\gamma_\mu>0$, the QNM eigenfunctions diverge as a further consequence of the radiation conditions; however, since $\mathbf{r}=\mathbf{r}_{\rm a}$ appears in the first part of $\Gamma^{\rm loss}$, while $\mathbf{r}\in V_{\rm gain}$ appears in $\Gamma^{\rm gain}$, the QNM expansion is well-defined for calculating the decay and pump rates of quantum emitters near cavity regions (cf. Ref.~\onlinecite{SI_short}). 
For the coupled disk resonator system, two hybridized QNMs $\mu=+,-$ appear in the optical frequency regime, which stem from the coupling of the fundamental modes of the isolated resonators~\cite{EPClassicalPaper}. Their respective field distributions is shown in Fig.~\ref{fig: Result1} (left), and they are used as the basis for the expansion of Eq.~\eqref{eq: GFQNM}.

We can now derive the Purcell factor $F_P$, i.e., the enhancement of the free-space emission rate $\Gamma_0$ of the TLS for the different models. Besides the LDOS-SE model, where the Purcell factor is defined as $F_P^{\rm LDOS}=\Gamma^{\rm LDOS}/\Gamma_0$ (using Eq.~\eqref{eq: FGR_class}), and the rigorous FGR model, where
$F_P^{\rm loss}=\Gamma^{\rm loss}/\Gamma_0$ (using Eq.~\eqref{eq: GammaLoss}), one can additionally define a Purcell factor from full numerical Maxwell-dipole calculations through Poynting vector terms~\cite{EPClassicalPaper},
which serves as a further calculation, that is independent from any mode expansion of the photon Green function. Here, $F_P=P/P_0$ and $P=\oint_\mathcal{S} \mathbf{S}\cdot \mathbf{n}$ is the power flow with the Poynting vector $\mathbf{S}={\rm Re}[\mathbf{E}\times\mathbf{H}]/2$ projected onto the normal vector $\mathbf{n}$ of a small spherical surface $\mathcal{S}$ enclosing a volume $\mathcal{V}$, that contains the dipole (this  does not intersect with the scattering structure, cf.~Fig.~\ref{fig: Result1}). Moreover, $P_0$ represents the power flow of the dipole without any scattering structures. Using Poynting's theorem, $P$ can be related to classical energy loss $P=-\int_\mathcal{V}{\rm Re}[\mathbf{j}^*\cdot\mathbf{E}]/2$, which is proportional to the dipole-projected LDOS.
Note that the LDOS and classical Purcell factors for an isolated gain resonator are
negative, but gain on its own does not constitute linear media.

The different model results as a function of TLS frequency are shown in Fig.~\ref{fig: Result1} (right). First, we note that the Purcell factor of the isolated lossy resonator disk (indicated by the grey dashed line) is roughly one order of magnitude smaller compared to the enhancements of the hybrid system, regardless of the model used. 
Second, and most significantly, the full numerical Maxwell-dipole solution as well as the result obtained from the LDOS-SE model show a highly non-Lorentzian lineshape, that produces {\em negative Purcell factors which are clearly unphysical}. Here, the gain-induced terms are taken into account as negative contributions, that compensate the loss. This conceptional problem is commonly ignored or never discussed in the literature, e.g., with  exceptional points connected to $\mathcal{PT}$-symmetric systems~\cite{Pick:17,PhysRevB.102.155303,PhysRevA.99.033853}.  
Note that although we show negative LDOS values, the two-QNM expansion is in excellent agreement with the full numerical Maxwell-dipole solution (without any approximations). Moreover, one should recognize that the power flow can indeed be negative in this situation, since the reference surface for calculating the enhancement does not enclose the whole resonator-emitter system. In contrast to the LDOS-SE model and the full numerical  Maxwell-dipole model, the results obtained from the corrected quantum model
are strictly positive.

\begin{figure}[ht]
   \includegraphics[trim={0.25cm 0.25cm 0.25cm 0.25cm},clip,width=1\columnwidth]{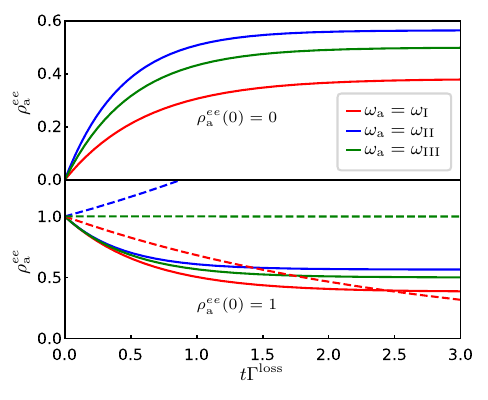}
   \caption{Temporal dynamics of the TLS excited state occupation $\rho_{\rm a}^{ee}$ with initial states $\rho_{\rm a}^{ee}(0)=0$ (top) and $\rho_{\rm a}^{ee}(0)=1$ (bottom) for the TLS frequencies indicated by vertical lines in Fig.~\ref{fig: Result1} using the full non-linear master equation (solid) 
   and the LDOS-SE model (dashed). Note, for $n_e(0)=0$, the full numerical Maxwell-dipole simulations and the LDOS-SE model predict $n_e(t)=0$, which is not shown here.}
   \label{fig: Result2}
\end{figure}

Next, we discuss the consequences of these models on the TLS population dynamics. 
We concentrate on three critical TLS frequency points that are indicated by vertical dashed lines in Fig.~\ref{fig: Result1}: (i) $\omega_{\rm a}=\omega_{\rm I}$, where $\Gamma^{\rm LDOS}$ reaches its maximum; (ii) $\omega_{\rm a}=\omega_{\rm II}$, where $\Gamma^{\rm loss}$ reaches its maximum (and $\Gamma^{\rm LDOS}$ its minimum), and (iii) $\omega_{\rm a}=\omega_{\rm III}$, where $\Gamma^{\rm LDOS}=0$. In the corrected quantum model, the temporal evolution of the densities and the dephasing is derived via Eqs.~\eqref{eq: rhoDynamics}(a-b), while in the LDOS-SE model, $\Gamma^{\rm loss}\rightarrow \Gamma^{\rm LDOS}$ and $\Gamma^{\rm gain}\rightarrow 0$. 
The results are shown in Fig.~\ref{fig: Result2}.  In the case of $\rho_{\rm a}(0)=|g\rangle\langle g|$ (TLS in the ground state initially), the corrected quantum model predicts a non-trivial time evolution, where the TLS occupation reaches a non-zero steady state with occupation $\rho_{\rm a}^{ee}(t\rightarrow\infty)=\Gamma^{\rm gain}/[\Gamma^{\rm loss}+\Gamma^{\rm gain}]$ after few decay processes ($\sim 3t\Gamma^{\rm loss}$), which is independent on the initial state. 
In contrast, the LDOS-SE 
model obviously predicts a trivial time evolution, since both the dephasing and the densities are proportional to the initial state. In the case of $\rho_{\rm a}(0)=|e\rangle\langle e|$ (TLS in the excited state initially), the difference is even more pronounced: while the result from the corrected quantum model again reveal non-trivial but physical meaningful steady states, the LDOS-SE model predicts unphysical solutions, that can either lead to infinite, constant or vanishing excited state populations for $t\rightarrow\infty$. 
This a consequence of the {\em wrong} interpretation of loss and gain-induced rates. We note that, although the dynamical equations \eqref{eq: rhoDynamics}(a-b) are formally similar to typically laser rate equations with thermal excitation, the steady states and interpretation of the gain sources is fundamentally different; while in the here presented case, two different reservoirs (gain and loss) are coupled to the emitter, the pumping in the typical models comes through a finite thermal photon number from a lossy environment.

To conclude, we have introduced a quantum dynamical approach to rigorously account for gain and decay of a TLS in linear dielectric media. In this formulation, the amplification enters the model through a reversed Lindblad term in a master equation, which only involves positive quantum rates. This is in complete contrast to the usual applied SE models, where the LDOS can exhibit negative values,  explicitly shown  for a coupled gain-loss resonator. This not only leads to a 
{\em fix} of the usually adopted (LDOS) SE of a TLS in an amplifying and absorptive environment, but  also implies a TLS pumping process through the gain region; this is a consequence of the operator ordering, manifesting in 
clear (and unique) quantum phenomena from vacuum QED. These  results 
also open up new avenues into currently proposed quantum theories for exceptional points and $\mathcal{PT}$-symmetric like system~\cite{PhysRevA.100.062131,PhysRevA.101.013812}, and future extensions of the theory towards strong light-matter coupling.
At a fundamental level, they also show that SE enhancement is indeed a uniquely quantum mechanical process that cannot always be described classically from Poynting's theorem.

\textit{Acknowledgements}.
We  acknowledge funding from Queen's University,
the Canadian Foundation for Innovation, 
the Natural Sciences and Engineering Research Council of Canada, and CMC Microsystems for the provision of COMSOL Multiphysics.
We also acknowledge support from the Deutsche Forschungsgemeinschaft (DFG) through SFB 951 Project B12 (Project number 182087777)
and the 
Alexander von Humboldt Foundation through a Humboldt Research Award.
This project has also received
funding  from  the  European  Unions  Horizon  2020
research and innovation program under Grant Agreement
No. 734690 (SONAR).

\end{document}